# Identification of Polytypism and Their Dislocations in Bilayer MoS$_2$ Using Correlative Transmission Electron Microscopy and Raman Spectroscopy

Xin Zhou [a,*], Tobias Dierke [b], Mingjian Wu [a], Shengbo You [a], Klaus Götz [c], Tobias Unruh [c], Philipp Pelz [a], Johannes Will [a], Janina Maultzsch [b], Erdmann Spiecker [a,*]

[a] Institute of Micro- and Nanostructure Research & Center for Nanoanalysis and Electron Microscopy (CENEM), Department of Materials Science & Engineering, Friedrich-Alexander-Universität Erlangen-Nürnberg, IZNF, Cauerstraße 3, 91058 Erlangen, Germany

[b] Department of Physics, Chair of Experimental Physics, Friedrich-Alexander-Universität Erlangen-Nürnberg, 91058 Erlangen, Germany

[c] Institute for Crystallography and Structural Physics (ICSP), Friedrich-Alexander-Universität Erlangen-Nürnberg, Staudtstraße 3, 91058 Erlangen, Germany

[*] Corresponding emails: xin.zhou@fau.de; erdmann.spiecker@fau.de;

Stacking orders and topological defects substantially influence the physical properties of 2D van der Waals (vdW) materials. However, the inherent features of 2D materials challenge the effectiveness of single characterization techniques in identifying stacking sequences, necessitating correlative approaches. Using bilayer MoS$_2$ as a benchmark, we differentiate its polytypism and specific dislocations through transmission electron microscopy (TEM) and Raman spectroscopy. Perfect and partial dislocations were revealed in TEM, which are closely linked to the stacking sequences, thus indirectly indicating the 2H and 3R polytypes. 3D electron diffraction reconstruction on relrods and low-frequency Raman spectroscopy further validated these polytypes owing to their reliance on crystal symmetry. Surprisingly, we unexpectedly resolved both polytypes despite starting with 2H bulk crystal, pointing to a possible phase transition during mechanical exfoliation. The correlative TEM-Raman approach can be extended to other 2D materials, paving the way for property alteration via stacking and defect engineering.

## Introduction

Atomically thin 2D vdW materials have garnered enormous interest since the groundbreaking work on monolayer graphene prepared by mechanical exfoliation, as they not only exhibit distinct properties but also open up a variety of unprecedented functionalities compared to their bulk counterparts. [1-3] Notably, these properties are strongly dependent on stacking order, symmetry, and the presence of defects. $MoS_2$, a prominent member of the 2D vdWs material family, exhibits multiple polytypes, among which the 2H and 3R phases represent the common stacking orders observed in natural and synthetic $MoS_2$. [2,4,5] These two phases behave considerably differently because of variations in stacking order and symmetry. For instance, 3R bilayer $MoS_2$ offers significant potential for piezoelectricity and spin/valley polarization owing to the broken inversion symmetry, while these properties are largely absent in 2H bilayer $MoS_2$ due to its retained inversion symmetry. [6-9] Dislocation, a sort of fundamental topological defect, can readily arise in vdW materials through interlayer sliding due to the inherent weakness of the interlayer vdW forces. The characteristics of dislocation are closely tied to the local atomic stacking configurations. Furthermore, dislocations are increasingly recognized as crucial elements in tuning materials' properties, such as plasticity and linear magnetoresistance. [2,10,11] Beyond the properties of intrinsic materials, assembling different layers with a controlled interlayer twist angle provides an exotic platform for exploring fruitful fundamental and novel phenomena in 2D materials. The local stacking orders of the twist-induced moiré superlattices and the screw dislocations describe the commensurate-incommensurate phase transitions, are heavily dependent on both the interlayer twist angle and the stacking configuration – either parallel (3R) or antiparallel (2H) stacking. [12,13] The remarkable effect of local stacking arrangement on the electronic structure of twisted bilayers has been extensively addressed, underscoring the potential applications of these materials in the rising field of twistronics. [14]

Therefore, the reliable identification of the various polytypes and the associated defects is a crucial starting step in comprehending and engineering the structure-dependent properties in 2D vdW materials. Though a wide range of measurement techniques have been developed, the reduced dimensions - being only a few atoms thick, the minute scale of topological defects, the strain and contamination residue in 2D vdW materials require the development of characterization techniques with nanoscale resolution and sensitivity to stacking sequence/symmetry. These stringent requirements limit the effectiveness of individual characterization techniques, emphasizing the necessity for combining different strategies. X-ray diffraction (XRD) is probably one of the most suitable techniques to investigate the crystal structure and quality of materials, but it suffers from complicated data interpretation owing to the atomically thin thickness and the interaction of X-rays with surrounding materials. [15,16] A broad family of scanning probe microscopy methods can identify local defects (such as vacancies), grain boundaries and qualify their physical properties. However, they are considerably surface- and strain-sensitive and can be influenced by layer number. [17,18] Optical spectroscopic approaches are promising for wafer-scale inspection and allow nondestructive investigation of the structural characteristics. For instance, Raman spectroscopy detects low-frequency interlayer modes (rigid-layer vibrations) and is useful in determining stacking sequence. [19,20] Yet, these methods are also susceptible to strain-induced effects [21,22] Most importantly, their spatial resolution is limited by the optical diffraction limit to roughly half a micrometer. This leads to great difficulties in discerning the defects in 2D materials, unless advanced methods like tip-enhanced spectroscopy is applied. [23,24] TEM excels with its imaging and diffraction capabilities, allowing for the direct observation of stacking orders and defects at the atomic level, [8,25] probing the crystal structure and symmetry. [26,27] However, TEM is often limited by a small field of view (typically in the micrometer range or less) and local buckling, making it difficult to provide a comprehensive overview of the material. Consequently, it is essential to integrate the strengths of multiple techniques and develop a correlative strategy to achieve complementary information about local atomic structure and defects in vdW materials before exploring their application across diverse fields further.

In this study, we use diffraction contrast and 3D electron diffraction (3D ED) to differentiate between 2H and 3R polytypes in exfoliated and freestanding bilayer $MoS_2$, along with dislocations present in the respective phases. We observed basal dislocations with distinct Burgers vectors across several bilayer regions, indicating the presence of different polytypes within the same flake. Quantitative evaluation of the relrod derived from 3D ED determines the stacking configuration unambiguously. The 2H and 3R polytypes are further confirmed by correlative low-frequency Raman spectroscopy conducted on the same sample sites. Our findings demonstrate an intuitive and scalable approach for determining stacking order in transition metal dichalcogenides (TMDCs), offering great potential for engineering their properties and applications through phase or defect modification.

## Results and discussion

We employed a step-by-step approach to prepare and transfer bilayer $MoS_2$ flake onto a TEM grid, as illustrated in Figure 1. The process starts with the mechanical exfoliation of a bulk $MoS_2$ single crystal using a transparent polydimethylsiloxane (PDMS) gel (Figure 1a, step 1). [28] Individual flakes were identified by light microscopy and transferred onto a 5 mm² Si substrate with a 300 nm $SiO_2$ layer. The $MoS_2$ flake produced in this way has a lateral size of up to 180 µm, according to the light microscopy image. The specific color and contrast of the flake against the $SiO_2$/Si substrate background offer a straightforward way to determine the number of $MoS_2$ layers annotated in the image. The layer numbers were confirmed afterwards by Raman- and photoluminescence (PL) spectroscopy, see Figure S8. The 300 nm $SiO_2$ layer displays a purple-colored background, and $MoS_2$ regions with varying numbers of layers exhibit different shades of blue. [29,30] In some cases, exfoliated flakes were cut by a laser to separate the bilayer part from the thicker part of the flakes, simplifying the transfer of the target flake onto a TEM grid (Figure 1b, step 2). QUANTIFOIL gold grid with an R1.2/1.3 holey carbon film was coated with a thin gold layer (see Experimental methods) to enhance the transfer efficiency, taking advantage of the strong affinity between gold and sulfur. [31] The grid was carefully aligned with the target flake under the light microscope, with the gold-coated side pointing towards the flake. A small droplet of isopropanol (IPA) was applied, followed by a small droplet of KOH solution (wt. 3%) after the IPA dried. The KOH solution gradually etched away the $SiO_2$/Si. After several minutes, the target flake was detached from the $SiO_2$/Si substrate and transferred onto the TEM grid (Figure 1c, step 3). Prior to TEM and Raman spectroscopy investigation, the sample was cleaned with water and an acetone bath, followed by heating on a hot plate (100 °C) for 10 minutes. The example illustrated in Figure 1 shows a large $MoS_2$ flake that spans two square windows of the TEM grid. TEM, Raman and PL studies were conducted on the same regions of the QUANTIFOIL grid, corresponding to the two marked bilayer regions.

In 2H and 3R $MoS_2$, the Mo atoms are sandwiched between two S layers in trigonal prismatic coordination, the S-Mo-S prisms of the two layers can be oriented either parallel or antiparallel, i.e., rotated by 180°. In the latter configuration, AA' also denoted as 2H stacking (Figure 2a), is the favored stacking arrangement with the lowest ground-state total energy. [2] In the case of parallel orientation of the prisms, AB (or AC) stacking is the preferred stacking order, with a ground-state total energy only slightly higher than 2H. [2] AB and AC are energetically equivalent and are commonly referred to as 3R stacking (Figure 2d), because the layer sequence is part of the ABC (or twinned ACB) stacking of bulk 3R. There are other possible stacking orders with high symmetry (AA, A'B, AB') but their respective ground-state total energies are considerably higher, in particular those of AA and A'B, where the chalcogen atoms of the two layers directly sit on top of each other. [2,32] Such stackings are not expected to be formed on large areas upon exfoliation, but may appear locally in defects, moiré lattices, or boundaries. [14,33-35]

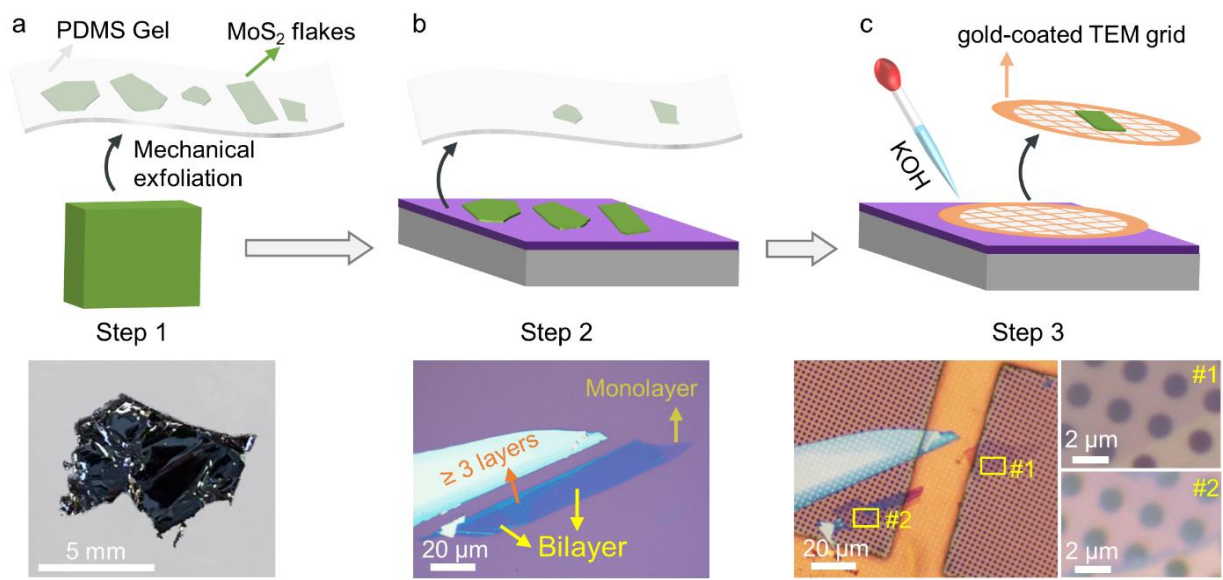

**Figure 1. Illustration of the step-by-step approach to prepare and transfer bilayer MoS₂ flake onto a gold-coated TEM grid**. (a and b) MoS₂ flakes were exfoliated using PDMS gel and transferred onto a SiO₂/Si substrate (steps 1 and 2). (c) The gold-coated TEM grid was carefully positioned in relation to the target flakes under the light microscope. The target flake was then transferred to the grid by sequentially applying IPA and KOH solution. Finally, the MoS₂ flake was cleaned using water and acetone bath, followed by heating on a hotplate in air.

Basal-plane dislocations can easily be introduced into the van der Waals gap of bilayer MoS₂ due to the weak interlayer bonding. However, the different symmetry and stacking order of 2H and 3R polytypes result in distinct dislocation characteristics. The main characteristic of a dislocation is its Burgers vector **b**, which describes the total displacement of one layer with respect to the other upon crossing the dislocation. The expected Burgers vectors for 2H and 3R MoS₂ are indicated by a red and blue arrow in Figures 2a and d, respectively. For the 2H phase, the Burgers vector is of type **b** = $(1/3)\langle\bar{2}110\rangle$, which corresponds to a lattice translation vector along a zigzag direction of the crystal. It is therefore a perfect dislocation, which means that the bilayer crystal has the same stacking on both sides of the dislocation. In contrast, the Burgers vector of a dislocation in 3R bilayer MoS₂ is of type **b** = $(1/3)\langle 0\bar{1}10\rangle$, oriented along one of the armchair directions of the crystal and is shorter than a lattice translation. Upon crossing the dislocation, the stacking order changes from AB to AC (or vice versa). In dislocation theory, such a dislocation is known as a partial dislocation, but it has also been referred to as a stacking boundary in the 2D materials community. [36,37] The reason why no perfect but only partial dislocations are expected in 3R bilayer MoS₂ is that the two stacking orders, AB and AC, are energetically equivalent, as has been discussed in detail for bilayer graphene. [38] Line direction **u** is another characteristic of a dislocation. While the Burgers vector of a dislocation is fixed, the line direction can change along the dislocation, giving the dislocation a screw (**u**||**b**), mixed (**u** inclined to **b**), or edge (**u** ⊥ **b**) character. Screw dislocations play a particular role in 2D moiré lattices, as they signal the formation of domain boundaries and introduce localized distortions, which ultimately affect the materials' physical properties. [12,14]

Figure 2 summarizes the investigation of isolated dislocations in the exfoliated bilayer $MoS_2$ using centered dark-field TEM (DF TEM) imaging and careful Burgers vector analysis. The stacking orders of the two regions in the bilayer $MoS_2$ were unknown at the time of the investigation but will be confirmed later through correlative electron diffraction, Raman, and PL spectroscopy. Figures 2b and c depict a series of $\{\bar{1}100\}$ and $\{\bar{2}110\}$ DF TEM images, revealing the dislocation in bilayer region #2 (Figure 1c). The corresponding diffraction pattern is shown in Figure S4 (Supplementary information) and the diffraction vector **g** used for dark-field imaging is indicated in each DF TEM image. The invisibility of the dislocation in the $(\bar{1}100)$ DF image suggests that the Burgers vector **b** is perpendicular to the diffraction vector **g** (according to the **g**·**b** = 0 invisibility criterion [11,38]) and thus parallel to $[\bar{1}\bar{1}20]$, as indicated by the red arrow. The sign of the Burgers vector is determined by investigating the dislocation contrast variation under different tilting angles (see details in Figure S5). The only crystallographically meaningful Burgers vector in this direction is **b** = (1/3) $[\bar{1}\bar{1}20]$, confirming a perfect dislocation as discussed above. These findings also (indirectly) indicate a 2H stacking of the bilayer $MoS_2$ in region #2. As expected, all other dark-field images in Figures 2b and c exhibit dislocation contrast, as no other **g** vector fulfills the invisibility criterion. Notably, the dislocation is seen with double line contrast in the $(\bar{1}\bar{1}20)$ DF image. This phenomenon is well known and results from the fact that |**g**·**b**| = 2 for this particular **g** vector. [39] In fact, the **g** vector is orientated parallel to the Burgers vector **b**, resulting in maximum scalar product **g**·**b** = (1/3)($\bar{1}\bar{1}20$) · $[\bar{1}\bar{1}20]$ = 2. For all the other **g** vectors, |**g**·**b**| = 1, in accordance with the observed single-line contrast. Observing that the $MoS_2$ exhibits identical contrast on both sides of the dislocation further supports the conclusion that the dislocation in region #2 is a perfect dislocation. The angle between the Burgers vector **b** and the dislocation line direction **u** indicates the dislocation has a mixed character with a larger edge component than the screw component. All evidences are pointing to the conclusion that the bilayer $MoS_2$ in region #2 has 2H stacking and that the dislocation is a perfect dislocation with Burgers vector **b** = (1/3) $[\bar{1}\bar{1}20]$ and mixed character.

Figures 2e and f present the corresponding DF TEM images and analysis for a dislocation located in region #1 (Figure 1c). The dislocation contrast is clearly different from that observed in region #2. The invisibility of the dislocation is achieved with the $(\bar{2}110)$ reflection, thus the Burgers vector of the dislocation is parallel to $[0\bar{1}10]$, as indicated by the blue arrow. The dislocation is almost a pure edge dislocation since the Burgers vector is approximately perpendicular to the dislocation line direction **u**. Taking crystallographic considerations into account again, the only meaningful Burgers vector is **b** = (1/3) $[0\bar{1}10]$, confirming the dislocation is a partial dislocation associated with 3R stacking. The stacking order changes from AB to AC (or vice versa) when moving from one side of the dislocation to the other. This change in stacking sequence is also consistent with the distinct contrast across the dislocation in the three $\{\bar{1}100\}$ DF TEM images (Figure 2e). The degree of contrast asymmetry depends on the exact orientation of the $MoS_2$ with respect to the incident electron beam. Even a slight deviation from the exact Bragg condition introduces a contrast difference that increases with increasing excitation error. [39] No contrast asymmetry is observed in the $\{\bar{2}110\}$ DF images (Figure 2f), as these reflections are insensitive to the change in stacking order from AB to AC. Apart from the $(\bar{2}110)$ DF image, all other DF images in Figures 2e and f show pronounced dislocation contrast, as expected for $\mathbf{g}\cdot\mathbf{b} \neq 0$. The dislocation contrast in the $(0\bar{1}10)$ DF image is particularly strong (and bright) compared to the dislocation contrast in the other two $\{\bar{1}100\}$ DF images. This is attributed to the fact that $\mathbf{g}$ is oriented parallel to $\mathbf{b}$, resulting in a larger magnitude of the scalar product $|\mathbf{g}\cdot\mathbf{b}| = 2/3$ than $|\mathbf{g}\cdot\mathbf{b}| = 1/3$ for the other two $\{\bar{1}100\}$ DF images. For the $(\bar{1}\bar{1}20)$ and $(1\bar{2}10)$ DF images, $|\mathbf{g}\cdot\mathbf{b}| = 1$, resulting in equally strong dislocation contrast. In summary, the DF TEM analysis in Figures 2e and f strongly suggests that the bilayer MoS₂ in region #1 has 3R stacking and that the dislocation is a partial dislocation with Burgers vector $\mathbf{b} = (1/3) [0\bar{1}10]$ and nearly pure edge character.

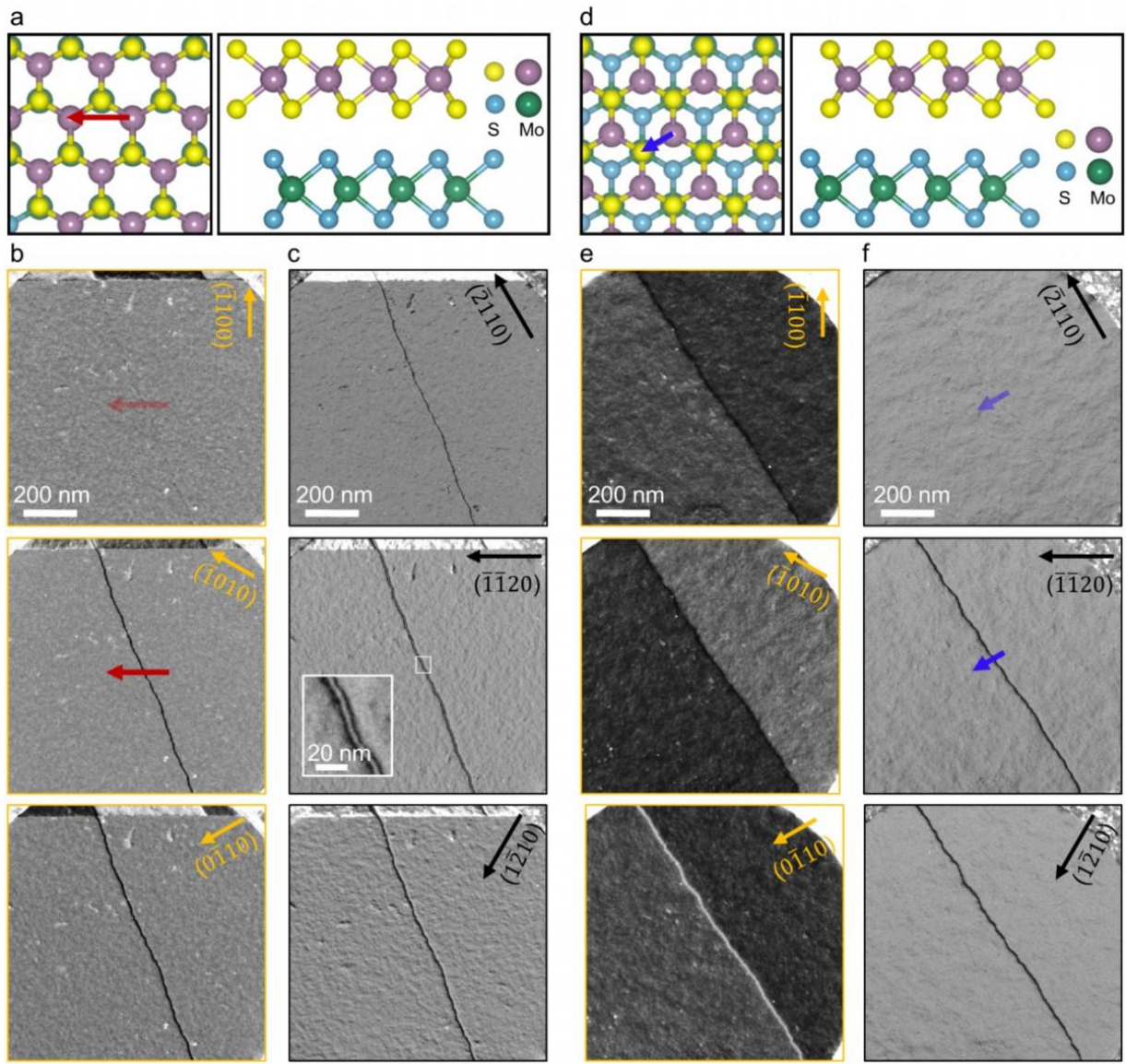

**Figure 2. Crystal structure and Burgers vector analysis on different types of dislocation, implying the presence of 2H and 3R polytypes in the exfoliated bilayer MoS₂ flake.** (a) and (d) Top and side views of crystalline structures for bilayer 2H and 3R MoS₂. The red and blue arrows in the top views represent the corresponding Burgers vectors of type $(1/3)\langle\bar{2}110\rangle$ and $(1/3)\langle 0\bar{1}10\rangle$. (b) and (c) Series of $\{\bar{1}100\}$ and $\{\bar{2}110\}$ DF TEM images acquired from the

bilayer region #2 in Figure 1, showing a perfect dislocation. (e) and (f) Series of $\{\bar{1}100\}$ and $\{\bar{2}110\}$ DF TEM images acquired from the bilayer region #1 in Figure 1, showing a nearly 90° partial dislocation.

The careful dislocation analysis described above provides indirect evidence of the stacking order - 2H in region #2 and 3R in region #1, through the specific dislocations associated with the respective bilayer $MoS_2$ polytype. To validate the stacking configuration and number of layers in the two sample regions, we performed correlative rocking curve analyses in TEM, followed by Raman and PL spectroscopy at the two marked regions in Figure 1c, which included the identical locations as the TEM investigation. We initially investigated the intensity oscillation of the relrods (rocking curves along $q_z$ direction) extracted from reconstructed 3D ED datacubes (complete 3D data visualizations are shown as insets in Figures 3a and 3b, and in Figure S6). These results are compared with simulated kinematic rocking curves for bilayer $MoS_2$ with various stacking sequences, including the low energy configurations of 2H and 3R phase, as well as other theoretically possible, high-energy configurations (Figure S7). [2,32] Two key pieces of information are yielded: I) layer number; II) symmetry and structure, reflected by first-order relrods. [40] As shown in Figures 3a and b, the experimental data fits the simulations well in terms of peaks positions and widths of the $\{1\bar{1}00\}$ and $\{11\bar{2}0\}$ relrods. The $\{11\bar{2}0\}$ relrods for both polytypes exhibit symmetric intensity oscillations and confirm the bilayer thickness. In contrast, the $\{1\bar{1}00\}$ intensity profiles vary dramatically between the 2H and 3R stacking configurations, owing to the differences in stacking sequence and crystal symmetry. For the intensity profile of $(1\bar{1}00)$ relrod in 2H stacking, the central peak nearly appears at the reciprocal lattice point $q_z$ = 0, but the whole profile exhibits a pronounced asymmetry, with a first intense peak on one side at -0.25 Å$^{-1}$ and a much smaller peak on the other side at 0.14 Å$^{-1}$ (red scattered dots and line in Figure 3a). For 3R bilayer, the $(1\bar{1}00)$ relrod intensity shows similar oscillations, but the central peak shifts to $q_z$ = -0.07 Å$^{-1}$. This shift originates from the displacement of the top layer towards the $[1\bar{1}00]$ crystalline orientation relative to the bottom layer. Additional simulated rocking curves for theoretically high-energy configurations (AB' and A'B stacking, Figure S7) reveal first-order relrod intensity profiles that clearly deviate from experimental observations. As a result, by comparing quantitative rocking curve evaluation to simulation, we unambiguously discriminate between the 2H and 3R polymorphs while ruling out additional high-energy configurations in bilayer $MoS_2$.

To further validate these findings on a large scale, Raman and PL measurements were conducted not only at the identical sample locations as in the TEM study but also at additional regions (the laser spots during Raman and PL measurements are indicated in Figure S8, determined by carefully aligning the TEM and light microscopy images). Figure 3c shows the Raman spectra of $MoS_2$ in the low-frequency range, with the shear mode (S) at a frequency of ~ 23 cm$^{-1}$ and the layer-breathing mode (LB) at ~ 40 cm$^{-1}$. The positions and shapes of the low-frequency modes are consistent with those of bilayer $MoS_2$. [41] Significantly, as shown in the top panel of Figure 3d, the amplitude ratio of LB/S in 2H stacking (~ 0.4) is smaller than that in 3R stacking (~ 1.4), due to the difference in the interlayer coupling between the two polytypes. A frequency shift in the S and LB modes from the 2H to the 3R polytype is also observed, leading to a slight increase in the frequency difference (~ 1.9 cm$^{-1}$) between the two modes (bottom panel in Figure 3d). Figures S8 c and d show additional Raman spectra of $MoS_2$ in the high-frequency region ($E_g$ and $A_{1g}$ modes), which further validate the bilayer thickness. The findings of the PL measurement results are displayed in Figure 3e, revealing two distinct peaks corresponding to the A and B excitons of bilayer $MoS_2$ with energies of ~ 1.81 eV and ~ 1.95 eV, respectively. The A and B exciton peak positions in 3R $MoS_2$ are red-shifted by 15.1 meV and 24.6 meV, respectively, compared to those in the 2H polytype, as seen in the inset of Figure 3e. Additionally, the B exciton appears more pronounced in the 3R structure. Both $MoS_2$ polytypes are therefore effectively distinguished using Raman and PL spectroscopy. In addition, comparative measurements reveal the splitting of $E_g$ mode in Raman spectra and considerable quenching in PL spectra for the measured regions with dislocations, regardless of stacking orders (Figures S8 c-d, g-h), indicating the presence of dislocations. However, it should be noted that not all holes containing dislocations exhibit such differences or signals in the Raman or PL spectra. Therefore, we believe a systematic investigation on the effects of dislocation on Raman and PL spectroscopy would be of great interest using tip-enhanced spectroscopy which exhibits spatial resolution on the scale of ~ 10 nm.

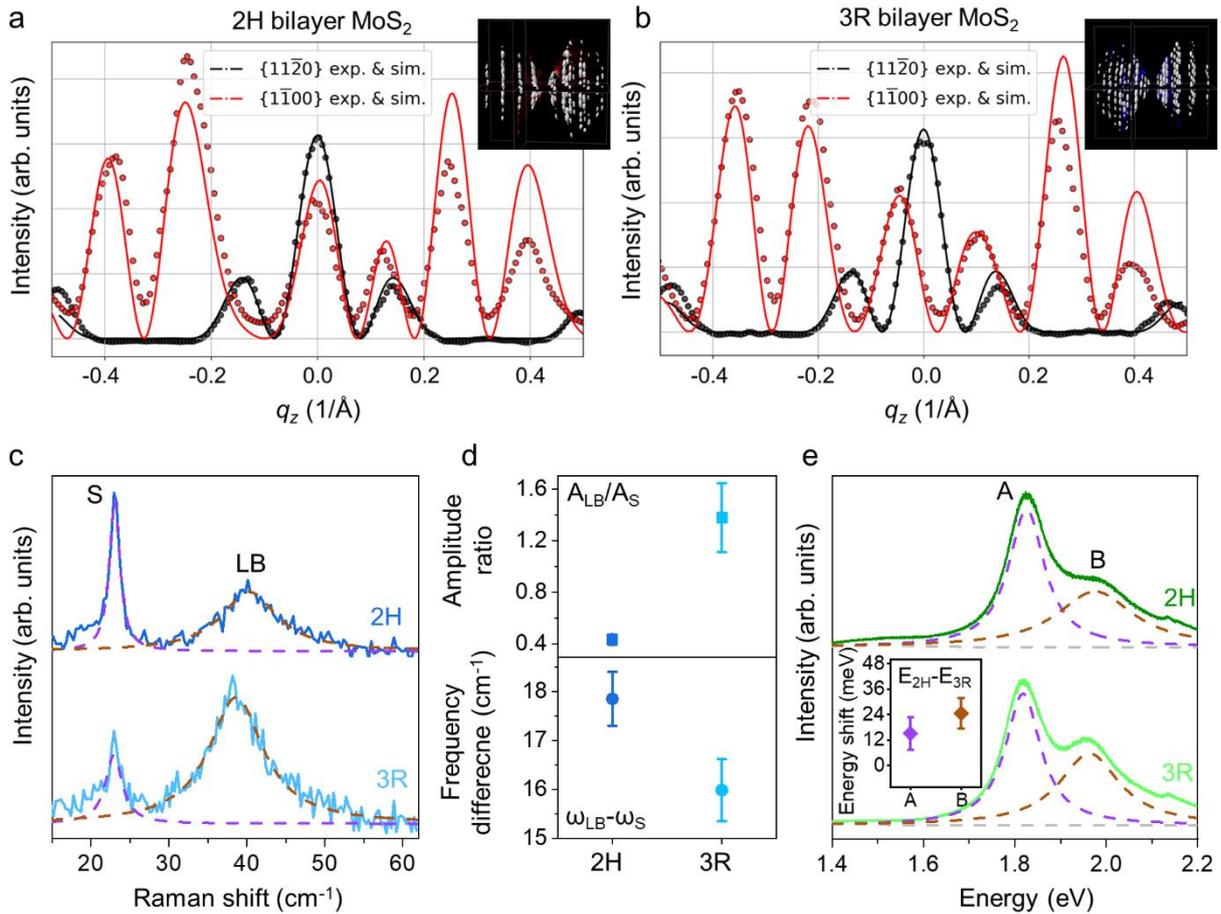

**Figure 3. Correlative determination of 2H and 3R polytypes within the bilayer $MoS_2$ flake using 3D ED, Raman and PL spectroscopy.** (a) and (b) Rocking curves of $(1\bar{1}00)$ (red dots) and $(11\bar{2}0)$ (black dots) reflections, extracted from 3D ED datasets (insets), which were acquired from dislocation-free areas of bilayer $MoS_2$. Solid lines are the simulated rocking curves. The rocking curves confirm the layer numbers as well as the different atomic staking orders (i.e., polymorphism). (c) and (e) Low-frequency Raman and PL spectra of the 2H and 3R bilayer $MoS_2$; the peaks were fit by Lorentzians (dashed lines). (d) The amplitude ratio and frequency difference between LB and S modes in 2H and 3R bilayer $MoS_2$. The inset in (e) highlights the average value of the A exciton and B exciton energy differences between 2H and 3R bilayer $MoS_2$. The error bars in (d) and inset in (e) represent the standard deviation derived from three independent Raman and PL spectra measured from three different sample regions (holes on the TEM grid).

## Conclusion

In summary, we have overcome the limitations of single-characterization techniques by implementing a correlative TEM-Raman approach to differentiate 2H and 3R polytypes in bilayer $MoS_2$. The different intensity oscillation and structural features in the first-order relrod, reconstructed through 3D ED construction, along with the variations in shear and layer breathing modes detected by Raman spectroscopy, allow for a comprehensive determination of the stacking orders in $MoS_2$. Furthermore, we identified perfect dislocation with Burgers vector of type **b** = $(1/3)\langle\bar{2}110\rangle$ in the 2H phase and partial dislocation with Burgers vector of type **b** = $(1/3)\langle0\bar{1}10\rangle$ in the 3R phase. These dislocations are closely related to their respective stacking arrangements. Interestingly, both polytypes were found within the same bilayer flake that was exfoliated from a parental 2H bulk crystal, implying a possible phase transition during exfoliation or transfer of the flake. Our study of stacking arrangements and dislocations in $MoS_2$ emphasizes the importance of correlative electron microscopy and optical-spectroscopy methods in investigating the structural and functional properties in 2D vdW materials.

## Experimental methods

### Sputter deposition of gold film on TEM Grids

Standard QUANTIFOIL Holey carbon R1.2/1.3 foils on 200 mesh gold grid is fixed onto the customized holder and placed into the HEX thin film sputter coater (Korvus Technology), with the carbon support film facing the gold source. After the evacuation of the sputter coater chamber, gold depositing starts along with the rotation of the grid (rotation speed, 5 rpm) to ensure uniform deposition. Eventually, a nominal thickness of 20 nm gold was deposited onto the gold TEM grids.

### Samples

Hexagonal $MoS_2$ bulk crystal was purchased from HQ Graphene. The 2H polytype was confirmed by X-ray diffraction, which showed well-developed pole figures of the 2H reflection $(103)_{2H}$, whereas no intensity was detected for the expected Bragg angle of the 3R reflection $(104)_{3R}$ (see Figure S1). Although this does not represent a full quantitative analysis of phase purity, the measurement demonstrates that the 2H phase is strongly dominant, and the presence of 3R stacking is negligible. Bilayer $MoS_2$ flakes were prepared by mechanical exfoliation of the bulk crystal using PDMS gel. Flakes with bilayer thickness were identified by light microscopy (LM, Zeiss Axio Scope) and then transferred to the gold TEM grid. Further details of the preparation procedure for correlative TEM, Raman and PL spectroscopy are described in the Results and discussion section.

### Transmission electron microscopy and diffraction rocking curves

TEM experiments were carried out using a Thermo Fisher Scientific (TFS) Spectra 200 microscope operated at 200 kV. The TEM is equipped with a cold field emission gun (X-CFEG), a Probe $C_s$-corrector, a high-resolution CMOS camera Ceta-S with ultra-sensitive scintillator (for TEM dark-field imaging), and a fast hybrid pixel detector Destris ARINA (for 4D scanning transmission electron microscopy, 4D STEM). For centered dark-field imaging (exposure time, 15 s), an objective aperture with size of 10 μm was used. The rocking curves were extracted from the 3D electron diffraction (3D ED) data, which were reconstructed based on nano-beam 4D STEM tilt series experiment. We have developed an automatic data acquisition workflow using SerialEM scripts [42] and SavvyScan external scan generator plug [43], along with a 3D ED data reconstruction routine based on simple polar transformation (see details in Supplementary information and Figure S2). Further experimental conditions include: Fishione Model 2021 analytical tomography holder, probe current of ~ 150 pA, calibrated convergence of 0.45 mrad (quasi-parallel illumination and the diffraction-limited probe size is estimated to be ~ 2.8 nm), camera length of 155 mm, dwell time of 100 μs, tilt range from -60° to 60° at 1° interval, which allows the rocking curve of $\{1\bar{1}00\}$ reflections to reach ± 0.6Å$^{-1}$ before cutting by the missing wedge. The spatially resolved 4D STEM datasets assure (via the virtual dark-field images) extracted diffraction patterns for 3D ED reconstruction are extracted from identical locations in each tilt angle, which can be critical given the known sample buckling. Alternatively, the rocking curves of selected reflection of interest were also acquired via dark-field imaging and tilt series experiment using a normal double tile holder (tilt range ± 25°). While the results are consistent, the limited tilt range is not sufficient to differentiate the theoretically possible high-energy stacking phases (see details in Figure S3).

Theoretical rocking curves for an equilibrium interlayer distance of 3.370 Å were calculated according to kinematic diffraction theory. We defined the lattice parameters as a = b = 3.161 Å, c = 12.295 Å for the 2H phase, and a = b = 3.163 Å, c = 18.37 Å for the 3R phase. [44] Supercells by extending c to be 200 Å, but containing only two layers of $MoS_2$ (thus only 6 atoms in the supercells) placed in the middle of the cells are constructed. Structure factor (F($\mathbf{q}$)),

$$F(\boldsymbol{q}) = \sum_i f_i(q) e^{2\pi i (\boldsymbol{q} \cdot \boldsymbol{r}_i)},$$

of the supercells are calculated using the ACOM module in the py4DSTEM package, where $\mathbf{q}$ is the scattering vector, $f_i$ is the atomic scattering factor, $r_i$ is the atomic position in the real space. [45,46] Since the bilayers are only 6 atoms thick, the diffracted intensity (I($\mathbf{q}$)) can be well described by kinematic diffraction theory, i.e.,

$$I(\boldsymbol{q}) \propto |F(\boldsymbol{q})|^2$$

The intensity of reflections of interest (i.e., {1100} and {2110}) are calculated by the square of the respective structure factors extracted from the full list of the calculated structure factor data. When comparing the experimental and simulation data, the intensity of the experimental data was scaled where the scaling factor is determined via least square fitting.

**Raman and Photoluminescence spectroscopy**

Raman and PL spectroscopy measurements on the bilayer $MoS_2$ flakes supported on the TEM grid were performed with a HORIBA LabRam HR Evolution spectrometer with laser excitation at λ = 532 nm and an 1800 lines/mm grating (300 lines/mm grating for PL measurements). The incident laser power was set to 30 μW to avoid heating effects and damage to the sample; the diameter of the laser spot on the sample was approximately 0.5 μm. The Raman and PL spectra were calibrated by neon lines. We performed thorough DF TEM investigation on the entire bilayer $MoS_2$ flake and carefully aligned the TEM images with the LM images, so that TEM, Raman, and PL results are correlatively collected from the identical measured regions.

**X-ray diffraction**

X-ray characterization of $MoS_2$ bulk flakes was conducted using a Rigaku SmarLab 9 kW diffractometer operating at 45 kV and 160 mA. A Johansson monochromator is utilized to monochromatized to the Cu-K$_{α1}$ wavelength of 1.54 Å. The beam is further parallelized to 0.05° and collimated to 5° by a Goebel mirror and Soller slits for Θ/2Θ measurements, respectively. For pole figure measurements the beam was further collimated towards 0.5°, which were then performed with a beam width and height of 2 and 1 mm, respectively.

## Acknowledgements

The authors gratefully acknowledge funding by the German Research Foundation (DFG) through the Collaborative Research Center SFB 953 "Synthetic carbon allotropes" and the Research Training Group GRK 1896 "In situ microscopy with electrons, x-rays and scanning probes." T.D. and J.M. acknowledge support by the Deutsche Forschungsgemeinschaft (DFG, German Research Foundation), project number 447264071 (INST 90/1183-1 FUGG).

## Competing interests

The authors declare no competing interests.

## Data Availability

The data that support the findings of our study are available upon request.